\documentclass[twocolumn,aps,pra,superscriptaddress]{revtex4}
\usepackage[latin9]{inputenc}
\usepackage{color}
\usepackage{amsmath}
\usepackage{amssymb}
\usepackage{graphicx}
\usepackage[unicode=true,
 bookmarks=false,
 breaklinks=false,pdfborder={0 0 1},backref=section,colorlinks=true]
 {hyperref}
\hypersetup{
 linkcolor=blue,citecolor=blue,urlcolor=blue}
\usepackage{breakurl}

\makeatletter
\@ifundefined{textcolor}{}
{%
 \definecolor{BLACK}{gray}{0}
 \definecolor{WHITE}{gray}{1}
 \definecolor{RED}{rgb}{1,0,0}
 \definecolor{GREEN}{rgb}{0,1,0}
 \definecolor{BLUE}{rgb}{0,0,1}
 \definecolor{CYAN}{cmyk}{1,0,0,0}
 \definecolor{MAGENTA}{cmyk}{0,1,0,0}
 \definecolor{YELLOW}{cmyk}{0,0,1,0}
}

\usepackage{bm}
\usepackage{color}
\usepackage{braket}
\usepackage{subfigure}\usepackage{epsfig}\usepackage{amsfonts}\usepackage{mathrsfs}\usepackage[toc,page,title,titletoc,header]{appendix}

\@ifundefined{textcolor}{}{%
 \definecolor{BLACK}{gray}{0}
 \definecolor{WHITE}{gray}{1}
 \definecolor{RED}{rgb}{1,0,0}
 \definecolor{GREEN}{rgb}{0,1,0}
 \definecolor{BLUE}{rgb}{0,0,1}
 \definecolor{CYAN}{cmyk}{1,0,0,0}
 \definecolor{MAGENTA}{cmyk}{0,1,0,0}
 \definecolor{YELLOW}{cmyk}{0,0,1,0}
}

\makeatother

\begin{document}

\title{Confinement induced resonance with weak bare interaction in a quasi 3+0 dimensional ultracold gas }

\author{Dawu Xiao}
\affiliation{Beijing Computational Science Research Center, Beijing, 100094, China}

\author{Ren Zhang}
\affiliation{School of Science, Xi'an Jiaotong University, Xi'an, 710049, China}

\author{Peng Zhang}
\email{pengzhang@ruc.edu.cn}
\affiliation{Beijing Computational Science Research Center, Beijing, 100094, China}
\affiliation{Department of Physics, Renmin University of China, Beijing, 100872,
China}

\begin{abstract}

  Confinement induced resonance (CIR) is a useful tool for the control of the interaction between ultracold atoms.  In most cases the CIR occurs when the characteristic length $a_{\rm trap}$ of the confinement is similar as the scattering length $a_{s}$ of the two atoms in the free three-dimensional (3D) space. If there is a CIR which can occur with weak bare interaction, i.e., under the condition $a_{\rm trap}\gg a_s$, then it can be realized for much more systems, even
  without the help of a magnetic Feshbach resonance, and would be very useful. In a previous research by P. Massignan and Y. Castin (Phys. Rev. A {\bf 74}, 013616 (2006)), it was shown that it is possible to realize such a CIR  in a quasi-(3+0)D system, where one ultracold atom is moving in the 3D space and another one is localized by a 3D harmonic trap. In this work we carefully investigate the properties of the CIRs in this system. We show that the CIR with $a_{\rm trap}\gg a_s$ can really occur, and the number of the CIRs of this type increases with the mass ratio between the moving and localized atoms.
  However,  when $a_{\rm trap}\gg a_s$ the CIR becomes extremely narrow, and thus are difficult to be controlled in realistic experiments.

\end{abstract}

\maketitle

\section{Introduction}
\label{intro}
In recent years, the ultracold atom gases becomes a very important platform for the experimental study of quantum many-body or few-body physics \cite{blochrmp}. One of the most important advantage of this system is that the interaction between two ultracold atoms, which is determined by the low-energy scattering amplitude, can be controlled via magnetic field or laser beam \cite{chinrmp}. In current experiments, the most widely-used approach for the control of inter-atomic interaction in the ultracold gases is the magnetic Feshbach resonance (MFR), with which one can tune the interaction via a bias magnetic field\cite{chinrmp,paulrmp}. So far many MFRs have been discovered and applied in the ultracold gases of various bosonic and fermionic atoms \cite{chinrmp}.

Nevertheless, the MFR technique still has some disadvantages. First,
in this approach the magnetic field should be fixed at a certain value, which is corresponding to the required inter-atomic interaction intensity. Thus, it is difficult to use the magnetic field to control other physical parameters or processes, {\it e.g.}, the coherence evolution of atomic spin states. Second, it is difficult to use the MFR for the control of spin-exchange interaction. This can be explained as follows. As mentioned before, to realized the MFR one needs to apply a bias magnetic field. As a result of this bias filed, there will be a Zeeman-energy gap between the two-atom hyperfine states before and after the spin-exchanging process, which is usually much larger than the kinetic energy of the relative motion of these two atoms. In this case, due to the energy conservation, the spin-exchanging process would be forbidden by this Zeeman-energy gap.

Therefore, to overcome these problems, one still needs to develop other approaches for the control of interaction between ultracold atoms.  One  option is the optical Feshbach resonance \cite{ofr1,ofr2,ofr3,ofr4,ofr5,ofr6,ofr7,ofr8,ofr9,ofr10}. In this approach, one can control the inter-atomic interaction via a laser beam which can induce two-body transitions when the two atoms are close to each other. However, a large heating effect can be induced by this laser beam, which significantly reduces the lifetime of the ultracold gas \cite{ofr1,ofr5,ofr6}.

Another important technique for the interaction-control of ultracold atoms is the confinement-induced resonance (CIR), with which one can tune the interaction via changing the size of the single-atom trapping potential. In realistic ultracold gases, the heating effect of the laser beam for the single-atom confinement is much weaker  than the one given by the laser beam for the two-body transition. Thus, the heating loss in a CIR is much weaker than the one in an optical Feshbach resonance. The CIR effect was discovered by M. Olshanii for quasi-one-dimensional (quasi-1D) ultracold gases \cite{cir1}. It has been theoretically discovered and studied \cite{cir1,cir2,cir3,cir4,cir7,Castin,ZP,cir11,cir12,cir13,cir14,m1,m2,m3,m4,m5,m6} for ultracold gases with various geometry of optical trapping, and generalized to  inelastic processes induced by the coupling between center-of-mass and relative motion \cite{cir9,cir10,cira,cirb} as well as other systems such as quantum dot \cite{cirdot} and atom-ion mixtures \cite{cir5,cirion,cirion2}. Moreover, CIR has also been realized in several experiments of ultracold gases  \cite{cir6,cir8,cirexp,cira}.

Nevertheless, the CIR technique also has one drawback. In most cases, a broad CIR can occur only when the scattering length $a_s$ of the two atoms in the 3D free space, which describes the intensity of the ``bare inter-atomic interaction", is of the same order of the characteristic length $a_{\rm trap}$ of the optical trapping potential. However, in most of the current experiments of ultracold gases, $a_{\rm trap}$ is at least of the order of $1000a_0$ (with $a_0$ being the Bohr's radius) while $|a_s|$ is at most of the order of $100a_0$. Therefore, to realize a CIR one requires to enhance $a_s$ via a MFR, and thus the disadvantages of the MFR which we introduced above still cannot be overcome.

Therefore, if a CIR  can appear under the condition $a_{\rm trap}\gg |a_s|$, it would be very useful. Using such a CIR one can control the inter-atomic interaction completely without the help of a MFR. As shown above, this would be very helpful for quantum simulations based on ultracold atoms.  In previous research two broad CIRs under the condition  $a_{\rm trap}\gg |a_s|$ for {\it negative} $a_s$ have been discovered \cite{Castin,cir11,cir12,cir13}. Nevertheless, for most types of ultracold atoms, the natural value of $a_s$ is {\it positive}. Some authors also found some CIRs which can occur for $a_{\rm trap}\gg a_s>0$ for some confinements. However, these CIRs are too narrow to be realized and controlled in experiments.

In 2006, P. Massignan and Y. Castin studied the CIR in a quasi (3+0)D system, where one atom (atom $A$) is in the 3D free space while the other one (atom $B$) is trapped by a 3D isotropic confinement~\cite{Castin}.  They considered the cases with mass ratio $m_A/m_B=0.15$, $m_A/m_B=1$ and $m_A/m_B=6.67$ (see Fig.~9$-$Fig.~11 of Ref.~\cite{Castin}), with $m_A$ and $m_B$ being the mass of atom A and B, respectively. It is shown that CIRs can occur when $a_{\rm trap}\gg a_s$ (e.g., $a_{\rm trap}/a_s=4$). More importantly, when the mass ratio $m_A/m_B$ is increased from 0.15 to 1 and to 6.67, it seems that the CIRs becomes more and more {\it broad}. This result implies the  perspective  that in this system with very high mass ratio (e.g., $m_A/m_B\approx 30$, like the mixture of $^{6}$Li and $^{171}$Yb atoms), the CIRs with $a_{\rm trap}\gg a_s$ can be broad enough, and then becomes applicable in current experiments.

In this work, we check this perspective via explicit calculations. We calculate the effective scattering length $a_{\rm eff}$ between these two atoms, which describes the intensity of the effective iner-atomic interaction and diverges at the CIR points, for mass ratio $m_A/m_B\leq 50$. For our system $a_{\rm eff}$ is a function of $a_{\rm trap}/a_s$.
Using our results we figure out the number and width of the CIRs as functions of the mass ratio.
Our results show that the number of the CIRs increase with $m_A/m_B$. On the other hand, however, the width of the CIRs increases with $m_A/m_B$ only when $m_A/m_B\lesssim 10$. For $m_A/m_B\gtrsim 10$, the CIR width in the ``$a_{\rm trap}/a_s$-axis" (with precise definition given in Sec. III) stops increasing  with $m_A/m_B$. Thus, there is an upper limit of the CIR width. Furthermore, for $a_{\rm trap}/a_s>3$ this upper limit is below $10^{-2}$. Namely, to realize and control the CIR in the experiment, the uncertainty of $a_{\rm trap}$ should be smaller than $10^{-2}a_s$. That is too difficult to be realized if  $a_s$ is not enhanced by a MFR. Thus, unfortunately, our results show that in this quasi (3+0)D system, the CIRs are still too narrow for the cases with $a_{\rm trap}\gg a_s$, and thus hard to be used for the control of interaction without an MFR. Nevertheless, our results are helpful for the understanding of two-body physics in the  quasi (3+0)D ultracold gases, which are important for the simulation of many important systems, e.g., the quantum open system and quantum gases with impurities. In addition, this work may also helpful for the searching of broad CIRs with weak background interaction (i.e., small $a_s$) in other systems.

The remainder of this paper is organized as follows. In Sec. II, we show our approach for the calculation of effective scattering length $a_{\rm eff}$. In Sec. III we illustrate our results and analysis the dependence of the number and width of the CIRs on the mass ratio. A summary and some discussions are given in Sec. IV. Some details of our calculations are shown in the appendix.

\section{Calculation of effective scattering length}
\label{sec1}
\begin{figure}[t]
\begin{center}
\includegraphics[width=5cm]{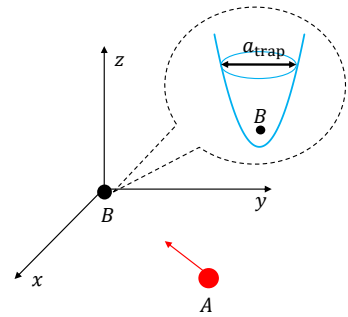}
\par\end{center}
\caption{(color online) Illustration of the quasi-(3+0)D scattering system. The heavy atom $A$ (incident atom) is moving in the 3D free space and the light atom $B$ is trapped by a 3D isotropic harmonic potential with characteristic length  $a_{\rm{trap}}$.}
\label{system}
\end{figure}

In this section, we show our approach for the calculation of the inter-atomic effective scattering length $a_{\rm eff}$, which is the method  in Ref.~\cite{Castin}. As shown in Fig.~\ref{system}, we consider the  scattering between two ultracold atoms $A$ and $B$, and assume that atom $A$ moves in the 3D space and atom $B$ is  trapped in a 3D isotropic harmonic potential. Thus, the Hamiltonian for our system is given by
\begin{eqnarray}
H=H_0+V,\label{h}
\end{eqnarray}
where $H_0$ is the two-body free Hamiltonian and can be written as ($\hbar=1$)
\begin{eqnarray}
H_{0}&=& -\frac{1}{2m_{A}}\nabla_{A}^{2}+H_B\label{h0}
\end{eqnarray}
with
\begin{eqnarray}
H_{B}&=&-\frac{1}{2m_{B}}\nabla_{B}^{2}+\frac{1}{2}m_{B}\omega^{2}{\bf r}_{B}^{2}.\label{hb}
\end{eqnarray}
Here ${\bf r}_{A(B)}$ and $m_{A(B)}$ are the coordinate and mass of atom $A(B)$, respectively, and $\omega$ is the frequency of the harmonic trapping potential for atom $B$, which is related to the characteristic length $a_{\rm trap}$ via
\begin{eqnarray}
a_{\rm trap}=\sqrt{\frac{1}{m_B\omega}}.\label{atrap}
\end{eqnarray}
We further define $\phi_{{\bf n}}\left({\bf r}_{B}\right)$  as
the eigen-state of the free Hamiltonian $H_B$ of the atom $B$,
with ${\bf n}=(n_x,n_y,n_z)$ being the corresponding quantum number in the $x$-, $y$- and $z$-directions. Explicitly, $\phi_{{\bf n}}\left({\bf r}_{B}\right)$ satisfies
\begin{eqnarray}
H_{B}\phi_{{\bf n}}\left({\bf r}_{B}\right)=E^{(B)}_{\bf n}\phi_{{\bf n}}\left({\bf r}_{B}\right),
\end{eqnarray}
where
\begin{eqnarray}
E^{(B)}_{\bf n}=\left(\frac{3}{2}+n_x+n_y+n_z\right)\omega
\end{eqnarray}
 being the corresponding eigen-energy of $H_B$.

In Eq. (\ref{h}), $V$ is the interaction potential between the ultracold atoms $A$ and $B$, which is modeled by the Huang-Yang pseudo potential
\begin{eqnarray}
V=\frac{2\pi a_{s}}{\mu_{AB}}\delta\left({\bf r}\right)\frac{\partial}{\partial r}\left(r\cdot\right),
\end{eqnarray}
where $\mu_{AB}$ is the reduced mass and ${\bf r}={\bf r}_A-{\bf r}_B $ is the relative position of these two atoms, and $a_s$ is the scattering length of these two atoms in the 3D space.

The effective scattering length $a_{\rm eff}$ is defined as
\begin{eqnarray}
a_{\rm eff}=-f_{{\bf k}={\bf 0}},\label{aef}
\end{eqnarray}
where $f_{{\bf k}={\bf 0}}$ is the scattering amplitude of these two atoms, corresponding to zero incident momentum of atom $A$. In the following, we solve this scattering problem and calculate $f_{{\bf k}={\bf 0}}$ and $a_{\rm eff}$. The incident state of our problem can be expressed as
\begin{eqnarray}
\psi^{\left(0\right)}\left({\bf r}_{A},{\bf r}_{B}\right)=\frac{1}{\left(2\pi\right)^{3/2}}\phi_{{\bf n}={\bf 0}}\left({\bf r}_{B}\right),\label{psi0}
\end{eqnarray}
where $\phi_{{\bf n}={\bf 0}}\left({\bf r}_{B}\right)$ is the ground state of $H_B$.
Notice that in Eq. (\ref{psi0}) the term $1/(2\pi)^{\frac 32}$ is nothing but the eigen-state of the momentum of atom $A$ with zero eigen-value, i.e., $e^{i{\bf k}\cdot{\bf r}_A}/(2\pi)^{\frac 32}|_{{\bf k}={\bf 0}}$.
It is clear that $\psi^{\left(0\right)}\left({\bf r}_{A},{\bf r}_{B}\right)$ is an eigen-state of the two-body free Hamiltonian $H_0$, with corresponding eigen-energy
\begin{eqnarray}
E_0=\frac{3}{2}\omega.
\end{eqnarray}
The scattering state $\psi^{\left(+\right)}\left({\bf r}_{A},{\bf r}_{B}\right)$ of our system satisfies the eigen-equation $H\psi^{\left(0\right)}\left({\bf r}_{A},{\bf r}_{B}\right)=E_0\psi^{\left(0\right)}\left({\bf r}_{A},{\bf r}_{B}\right)$,
as well as the out-going boundary condition for atom $A$ in the limit $r_A\rightarrow+\infty$. This equation and boundary condition can be re-formulated as the Lippman-Schwinger type equation
\begin{align}
&\psi^{\left(+\right)}\left({\bf r}_{A},{\bf r}_{B}\right)-\psi^{\left(0\right)}\left({\bf r}_{A},{\bf r}_{B}\right)\nonumber \\
=&\frac{2\pi a_s}{\mu_{AB}}\int  d{\bf R}G_{0}\left({\bf r}_{A},{\bf r}_{B};{\bf R},{\bf R}\right)\eta({\bf R}),
\label{eq:LPEquation}
\end{align}
where the function $\eta\left({\bf R}\right)$ is related to the wave function $\psi^{\left(+\right)}\left({\bf r}_{A},{\bf r}_{B}\right)$ via
\begin{equation}
\eta\left({\bf R}\right)=\left.\frac{\partial}{\partial r} \left[r\psi^{\left(+\right)}\left({\bf R}+\frac{m_{B}}{M}{\bf r},{\bf R}-\frac{m_{A}}{M}{\bf r}\right]\right)\right|_{r\rightarrow0}
\label{eta}
\end{equation}
with $M=m_A+m_B$. Here $G_0$ is the free Green's function which can be expressed as
\begin{eqnarray}
G_{0}\left({\bf r}_{A},{\bf r}_{B};{\bf r}'_{A},{\bf r}'_{B}\right)&=&\langle{\bf r}_{A},{\bf r}_{B}\vert\frac{1}{E_0+i0^{+}-H_{0}}\vert{\bf r}'_{A},{\bf r}'_{B}\rangle\nonumber\\
\end{eqnarray}
in terms of the Dirac bracket, where $|{\bf r}_A,{\bf r}_B\rangle$ is the eigen-state of the position operator of the two atoms. Then, it can be re-expressed as
\begin{align}
&G_{0}\left({\bf r}_{A},{\bf r}_{B};{\bf r}'_{A},{\bf r}'_{B}\right)\nonumber \\
= & -\frac{m_A \phi_{{\bf n}={\bf 0}}\left({\bf r}_{B}\right)\phi_{{\bf n}={\bf 0}}^{\ast}\left({\bf r}'_{B}\right)}{2\pi\left|{\bf r}_{A}-{\bf r}'_{A}\right|} \nonumber \\
& -\sum_{{\bf n}'\neq {\bf 0}}\frac{m_Ae^{-\kappa_{{\bf n}'}\left|{\bf r}_{A}-{\bf r}'_{A}\right|}\phi_{{\bf n}'}\left({\bf r}_{B}\right)\phi_{{\bf n}'}^{\ast}\left({\bf r}'_{B}\right)}{2\pi\left|{\bf r}_{A}-{\bf r}'_{A}\right|},\label{eq:G0_1}
\end{align}
with $\kappa_{{\bf n}'}=\sqrt{(2m_{A})\left(n'_{x}+n'_{y}+n'_{z}\right)\omega}$.
Therefore, in the long-range limit $r_A\rightarrow\infty$ the scattering wave function has the asymptotic expression
\begin{equation}
\psi^{\left(+\right)}\left({\bf r}_{A},{\bf r}_{B}\right)=\frac{1}{(2\pi)^{\frac32}}\left(1+f_{{\bf k}={\bf 0}}\frac{1}{r_{A}}\right)\phi_{0}\left({\bf r}_{B}\right)\label{psip}
\end{equation}
with the zero-momentum scattering amplitude $f_{{\bf k}={\bf 0}}$ being given by
\begin{equation}
f_{{\bf k}={\bf 0}}=-\frac{m_{A}}{2\pi}\int d{\bf R}\phi_{0}^{*}\left({\bf R}\right)\eta\left({\bf R}\right).\label{fk}
\end{equation}
Similar as above, in the Eq. (\ref{psip}) of the scattering state $\psi^{\left(+\right)}\left({\bf r}_{A},{\bf r}_{B}\right)$, the term ``1" and $1/r_A$ is just the incident plane wave and out-going spherical wave with zero momentum, respectively, i.e., $1=e^{i{\bf k}\cdot{\bf r}_A}|_{{\bf k}={\bf 0}}$ and $1/r_A=e^{ikr_{A}}/r_A|_{{\bf k}={\bf 0}}$.

According to Eq. (\ref{aef}) and Eq. (\ref{fk}), we know that the effective scattering length $a_{\rm eff}$ can be expressed as
\begin{equation}
a_{\rm eff}=\frac{m_{A}}{2\pi}\int d{\bf R}\phi_{0}^{*}\left({\bf R}\right)\eta\left({\bf R}\right).\label{aef2}
\end{equation}
On the other hand, as shown in the Appendix A, the function $\eta\left({\bf R}\right)$ defined in Eq. (\ref{eta}) satisfies an integral equation
\begin{equation}
\eta\left({\bf R}\right)=\psi^{(0)}({\bf R},{\bf R})+\hat{O}[\eta\left({\bf R}\right)],\label{ie}
\end{equation}
where $\hat{O}$  is an integral operator whose expression is given in Appendix A.

We numerically solve the integral equation Eq. (\ref{ie}) and derive the function $\eta\left({\bf R}\right)$. Then we substitute the result into Eq. (\ref{aef2}) and obtain the effective scattering length $a_{\rm eff}$.

\section{Properties of the CIRs}
\label{sec2}

In the above section we show the approach for the numerical calculation of the effective scattering length $a_{\rm eff}$ between the flying atom $A$ and the trapped atom $B$. In this section we show our numerical results and investigate the
properties of the CIRs in this system. As shown in Sec. I, we focus on the cases where the bare scattering length $a_s$ is {\it positive}, and the
mass of atom $A$ is much larger than the mass of atom $B$.

\begin{figure}[t]
\begin{center}
\includegraphics[width=7cm]{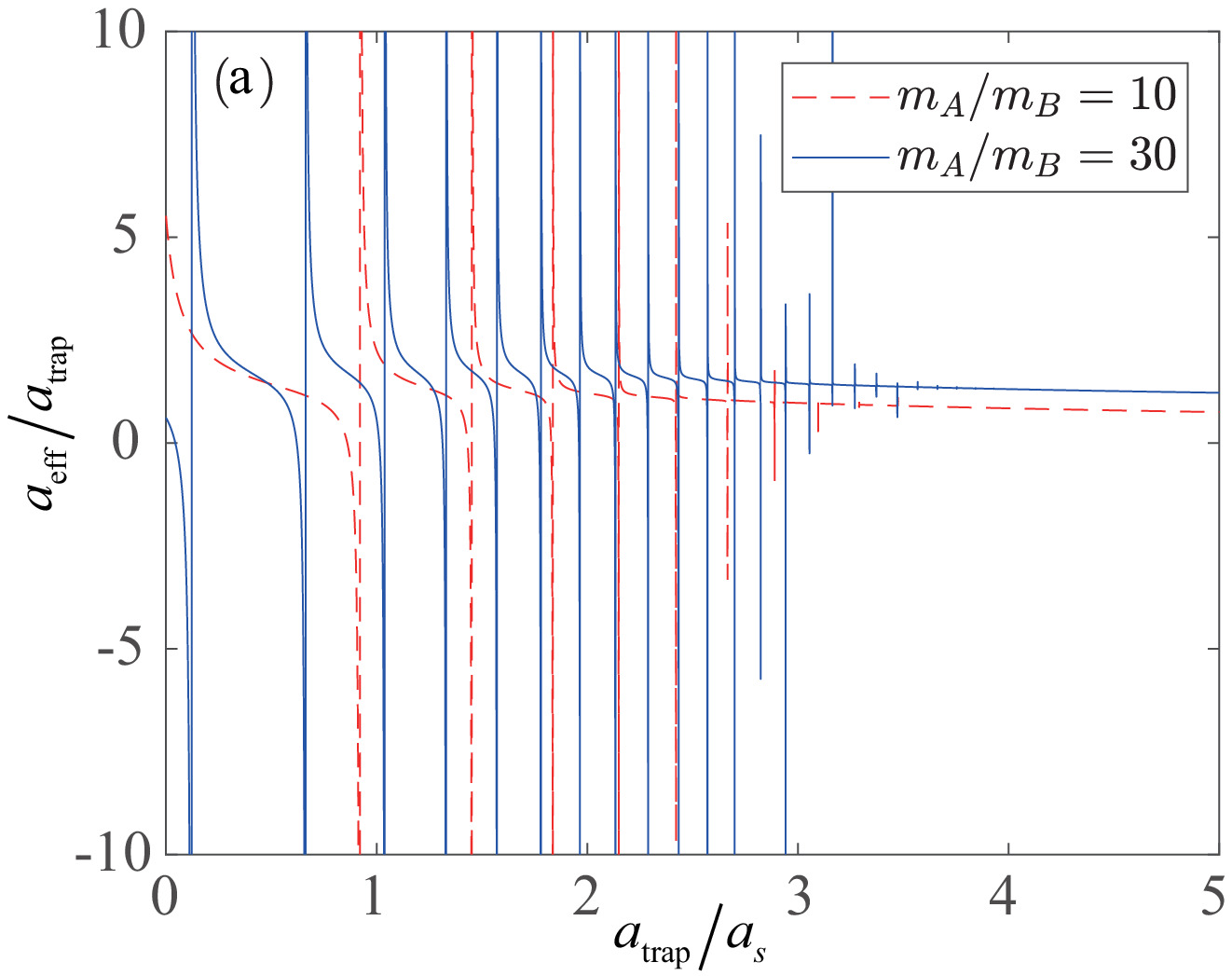}
\includegraphics[width=7cm]{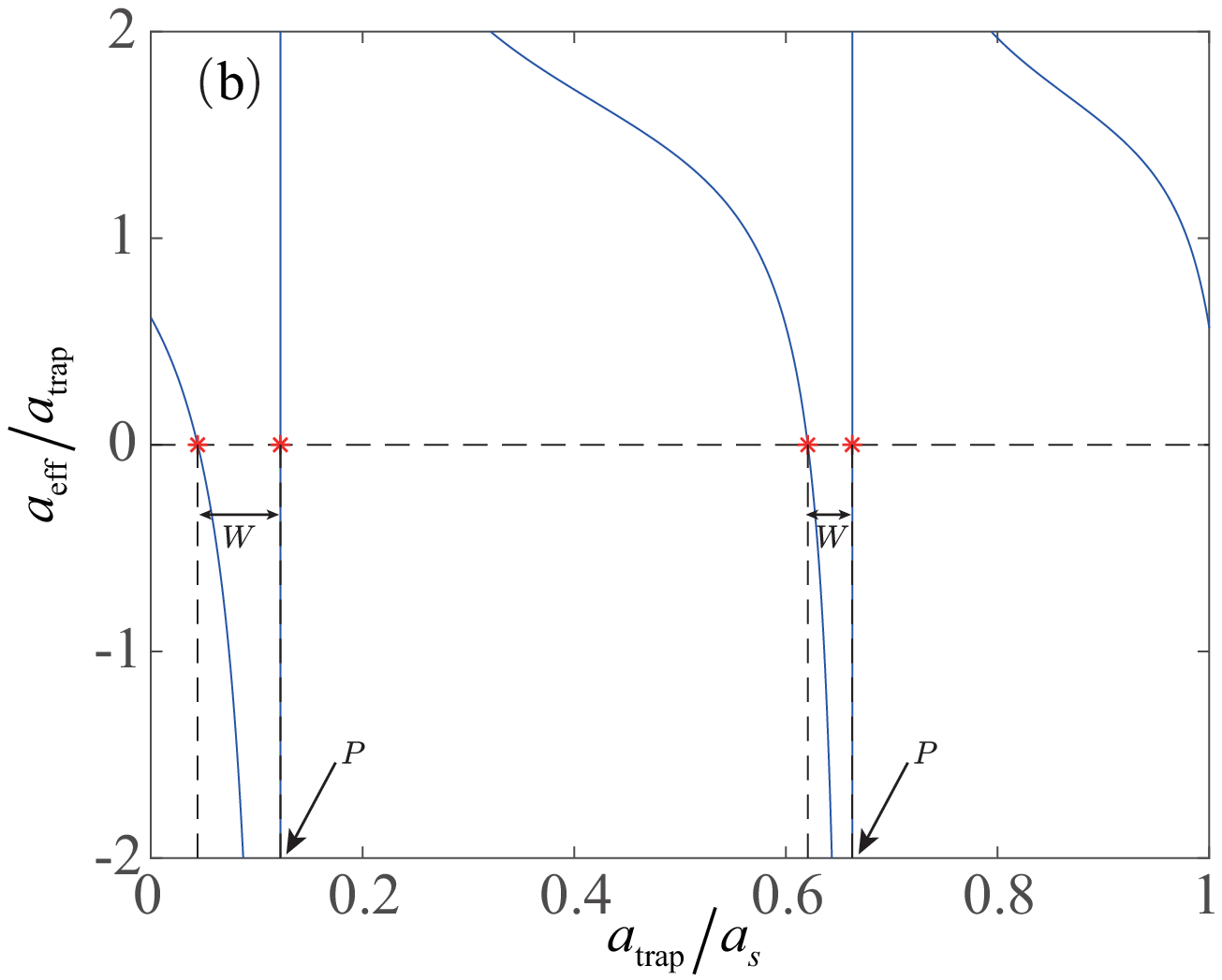}
\par\end{center}
\caption{(color online) {\bf (a):} The effective scattering length $a_{{\rm eff}}$ as a function
of $a_{\rm trap}/a_{s}$ for $m_A/m_B=10$ (red dashed line) and $m_A/m_B=30$ (blue solid line). {\bf (b):} Illustration of the definition
of the CIR point $P$ and CIR width $W$.}
\label{res}
\end{figure}

In Fig.~\ref{res}(a) we show the effective scattering length $a_{{\rm eff}}$ as a function of  $a_{\rm trap}/a_s$ for the cases with mass ratio $m_A/m_B=10$ and  $m_A/m_B=30$. As defined in Eq. (\ref{atrap}), $a_{\rm trap}$ is the characteristic length  of the confinement of atom $B$.
It is clearly shown that $a_{\rm eff}$ diverges when $a_{\rm trap}/a_s$ takes several certain values. Namely,
multi-CIRs can appear.

To further investigate the properties of these CIRs, here we define the position and width of each CIR. As shown in Fig.~\ref{res}(b), the position $P$ of a CIR is naturally defined as the divergent place of $a_{\rm eff}$, i.e., we have $a_{\rm eff}=\infty$ for  $a_{\rm trap}/a_s=P$. Furthermore, we define the width $W$ of a CIR as the distance between the CIR position $P$ and the nearest zero-crossing point of $a_{\rm eff}$.
As mentioned in Sec. I, $a_{\rm eff}$ describes the intensity of the effective interaction between the two atoms. Thus, at a CIR point $P$ the effective interaction is enhanced. In addition, the CIR width $W$ describes the broadness of the CIR, or the size of the parameter region where $a_{\rm eff}$ can be effectively controlled via the CIR. Explicitly, in  the region $a_{\rm trap}/a_s\in[P-W,P+W]$ one can efficiently tune $a_{\rm eff}$ by changing $a_{\rm trap}$.

Here we would like to emphasis that, for a system with fixed bare scattering length $a_s$, to precisely control the effective inter-atomic interaction (i.e., control $a_{\rm eff}$) by changing $a_{\rm trap}$ via a CIR, the uncertainty of $a_{\rm trap}$ in the experiment should be much smaller than $Wa_s$ of this CIR. For the cases where $a_s$ is not enhanced by a MFR and is only of the order of $100a_0$, which we are interested in, if $W\lesssim10^{-2}$,  the uncertainty of $a_{\rm trap}$ should be less than $1a_0$. That is difficult to be realized in the experiments. Therefore, a CIR which is helpful for the control of inter-atomic interaction should be broad enough.

\begin{figure}[t]
\begin{center}
\includegraphics[width=7cm]{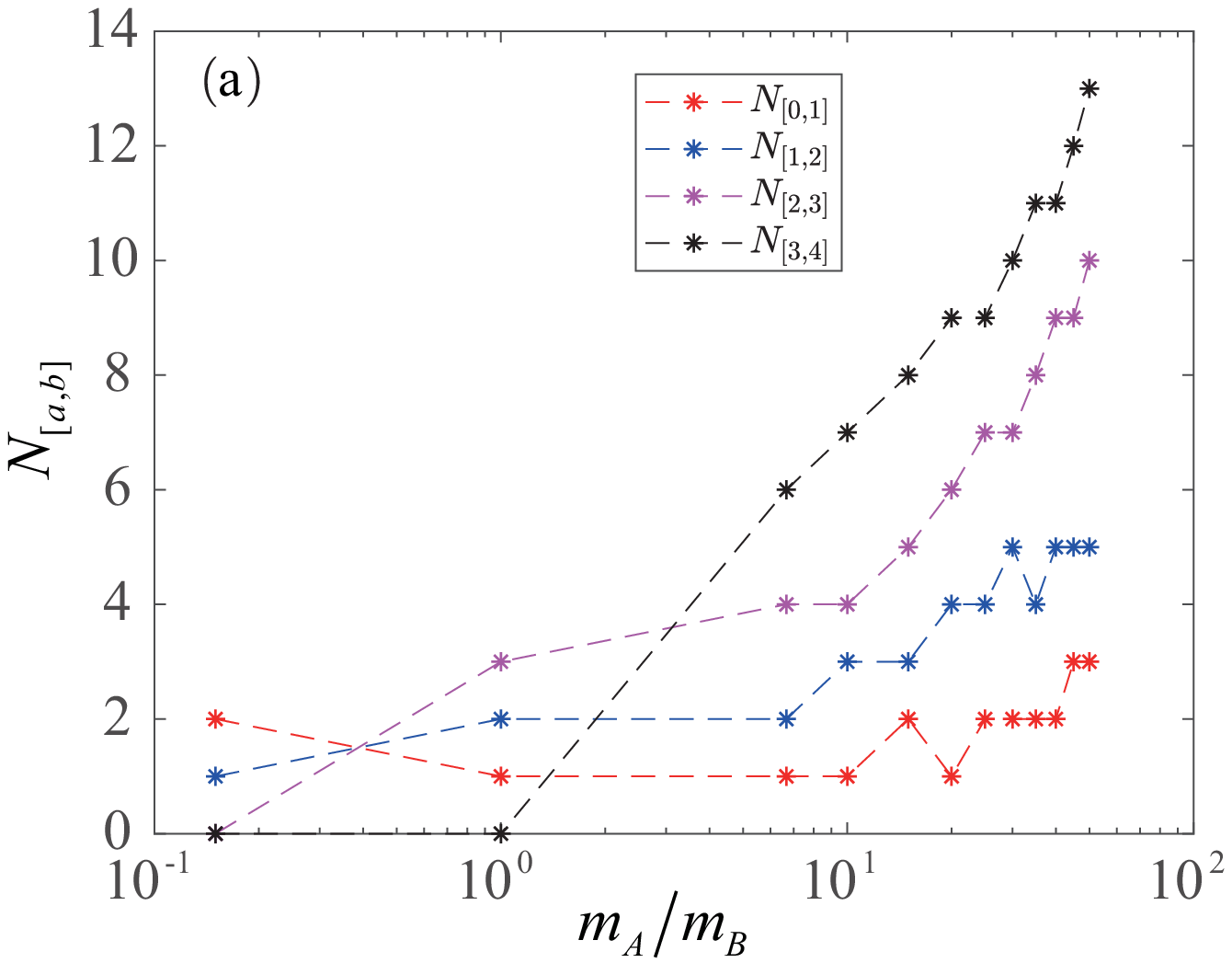}
\includegraphics[width=7cm]{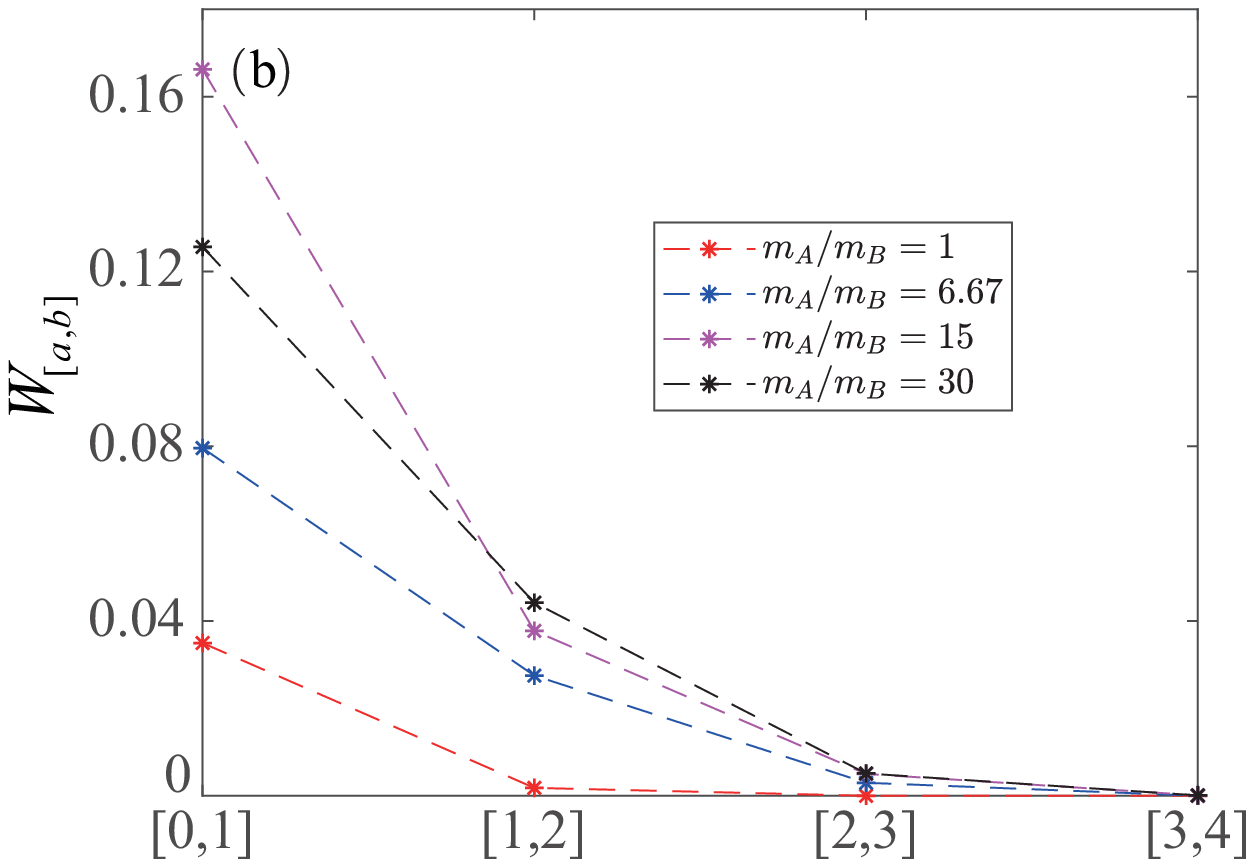}
\includegraphics[width=7cm]{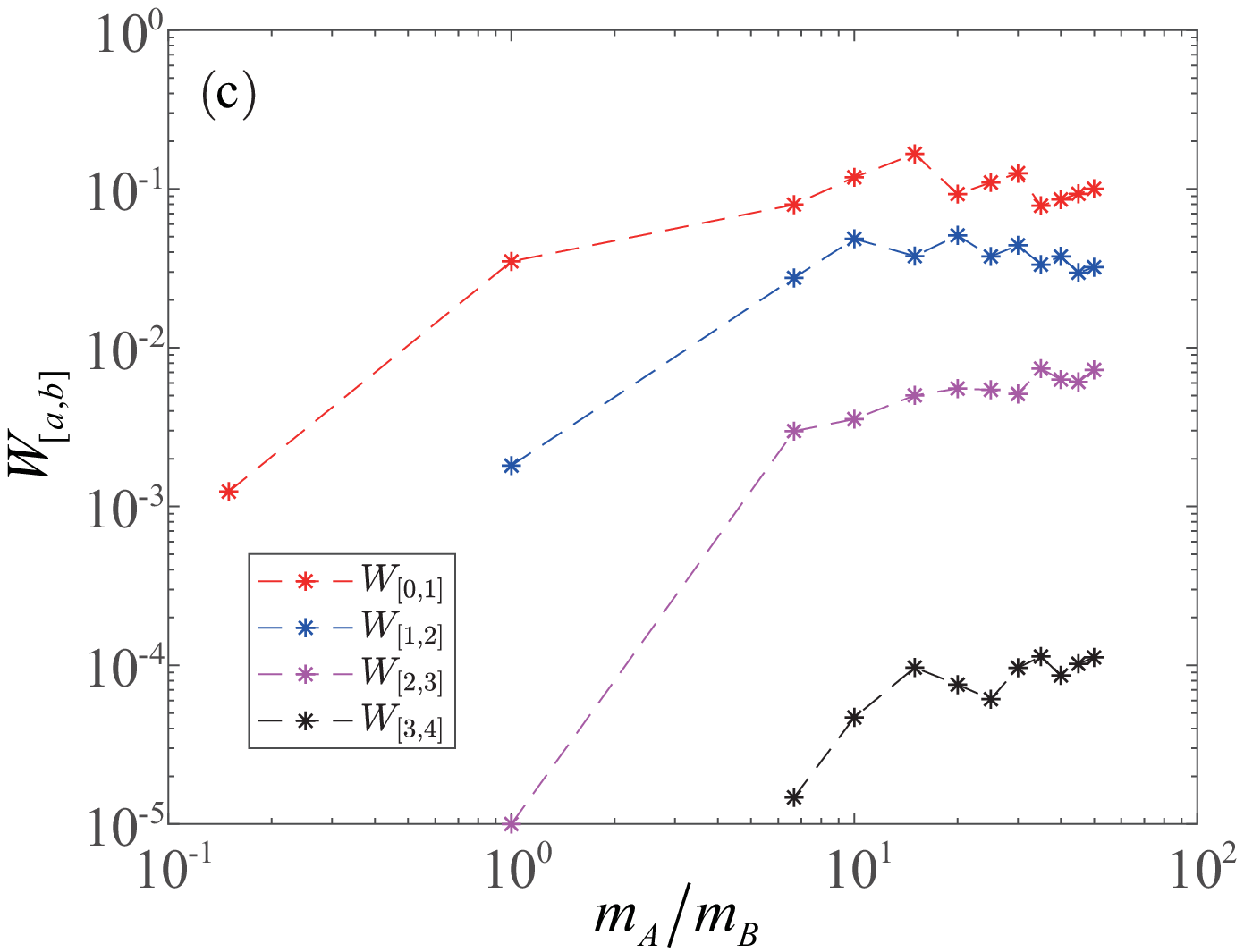}

\par\end{center}
\caption{(color online) {\bf (a):} The number of CIRs in the regions $a_{\rm trap}/a_s\in[0,1]$, $[1,2]$, $[2,3]$ and $[3,4]$, as functions of the mass ratio $m_A/m_B$.
{\bf (b):} The maximum resonance
width $W_{[a,b]}$ of CIRs in the regions $a_{\rm trap}/a_s\in[a,b]$ ($[a,b]=[1,2]$, $[2,3]$ and $[3,4]$) for different mass ratios.
{\bf (c):} $W_{[a,b]}$ as a function of the mass ratio $m_A/m_B$.}
\label{NW}
\end{figure}

We further define $N_{[a,b]}$ as the number of the relatively broad CIRs with $W>10^{-5}$ in the region $a_{\rm trap}/a_s\in[a,b]$. Here we emphasis that, our calculations show that in this system many  CIRs with $W<10^{-5}$ can appear. However, in this work we do not take into account these extremely-narrow CIRs, because (i) as shown above, these CIRs are too narrow and not feasible for the experiments, and (ii) these CIRs are also too narrow to be clearly resolved via our numerical calculations.

In Fig.~\ref{NW}(a) we show the CIR number $N_{[0,1]}$, $N_{[1,2]}$, $N_{[2,3]}$ and $N_{[3,4]}$ for the regions $a_{\rm trap}/a_s\in[0,1]$, $[1,2]$, $[2,3]$ and $[3,4]$, as functions of the mass ratio $m_A/m_B$. It is shown that the amount of the CIRs increases with $a_{\rm{trap}}/a_s $ and the mass ratio $m_A/m_B$. Namely, we can find more CIRs when the flying atom $A$ is more heavier than the trapped atom $B$, and the bare scattering length $a_s$ is more smaller than the size $a_{\rm trap}$ of the trap.

Furthermore, as shown in Sec. I, the main question of this work is: can we increase the width of the CIRs by increasing the mass ratio $m_A/m_B$, especially for the cases with small $a_s$? To answer this question, we need to study the variation of the width of the CIRs with the mass ratio $m_A/m_B$ and the ratio $a_{\rm trap}/a_s$. Nevertheless, since there are many CIRs in our system, it is difficult and not necessary to investigate the width of each one. For simplicity, here we introduce a parameter $W_{[a,b]}$ which is defined as the maximum value of the widths of the CIRs in the region $a_{\rm trap}/a_s\in[a,b]$. It is clear that  properties of the CIR width in our system can be described by the behavior of $W_{[a,b]}$.

 In Fig.~\ref{NW}(b, c) we show the maximum width $W_{[0,1]}$, $W_{[1,2]}$, $W_{[2,3]}$ and $W_{[3,4]}$ for various  mass ratio $m_A/m_B$. As shown in Fig.~\ref{NW}(b) that for a fixed mass ratio, the width of the CIRs {\it decreases} with the ratio $a_{\rm trap}/a_s$. Thus, the CIRs becomes narrower for smaller $a_s$. On the other hand, as shown in Fig.~\ref{NW}(c), for each given region of $a_{\rm trap}/a_s$, when the mass ratio $m_A/m_B\lesssim 10$, the increasing of CIR width increases with $m_A/m_B$. Namely, the CIRs becomes broader when the flying atom $A$ is heavier. That is consistent with the results shown in Fig. 9-11 of Ref.~\cite{Castin} for $m_A/m_B=0.15$, 1 and 6.67. However, when $m_A/m_B\gtrsim 10$, the increasing of the CIR width stops. Therefore,
 for a system with fixed $a_{\rm trap}$ and $a_s$, we cannot always increase the widths of the CIRs by increasing $m_A/m_B$. Namely,
the CIR width has an upper limit. Unfortunately, Fig.~\ref{NW}(c) also clearly shows that this upper limit of the CIR width {\it decreases} with $a_{\rm trap}/a_s$. For $a_{\rm trap}/a_s\in[2,3]$, this upper limit is already below $10^{-2}$. As shown above, this means that the CIRs are too narrow and not feasible for the experiments.
Therefore, for the case with $a_s\sim100a_0$ and $a_{\rm trap}\sim 1000a_0$, which we are interested in, it is hard to use the CIRs in this quasi-(3+0)D system to control the inter-atomic interaction in the experiments.

\section{Summary}
\label{Summary}

In this work, we investigate the properties of the CIRs of the scattering between the atom $A$ which is moving in the 3D space and the atom  $B$ which is localized by an harmonic trapping potential. We consider the cases with positive bare scattering length $a_s$ and various mass ratio $m_A/m_B$. We show that the amount of the CIRs increases with $m_A/m_B$. On the other hand, the width of the CIRs increases with $m_A/m_B$ for $m_A/m_B\lesssim 10$, and this increasing stops for $m_A/m_B$ for $m_A/m_B\gtrsim 10$. Furthermore, the upper limit of the CIR width decreases with the ratio $a_{\rm trap}/a_s$. Explicitly, for $a_{\rm trap}/a_s\gtrsim 3$ the widths of the CIRs are smaller than $10^{-2}$. Therefore, for these cases  the CIRs  are too narrow to be used for the control of inter-atomic interaction in experiments. Thus, if the bare scattering length $a_s$ is not enhanced by a MFR and only of the order of $100a_0$, and  $a_{\rm trap}$ is of the order of $1000a_0$ as in the current experiments, it is difficult to use CIR of this quasi (3+0)D system to control the inter-atomic interaction. Our results are helpful for understanding the two-body problem in this mixed-dimensional system, and for finding the CIRs which can occur for small 3D scattering length and are broad enough for experimental control in other confined ultracold gases.

\begin{acknowledgements}
This work is is supported in part by the National Key Research and Development Program
of China Grant No. 2018YFA0306502 (PZ), No. 2018YFA0307601(RZ), NSFC (Grant No. 11434011(PZ), Grant No. 11674393 (PZ), Grant No.11804268 (RZ), Grant No. 11534002),
NSAF (Grant No. U1530401 and Grant No. U1730449), as well as the Research Funds of Renmin University
of China under Grant No. 16XNLQ03(PZ).
\end{acknowledgements}

\newpage
\begin{widetext}
\appendix

\section{Integral equation}
\label{Appendix}

In this part, we derive the integral equation of the function $\eta\left({\bf R}\right)$ defined in Sec. II.
Here we also emphasis that our result is the same as Ref.~\cite{Castin}, while our
method  is slightly different with the method of Ref.~\cite{Castin}. Explicitly, in Ref.~\cite{Castin} the integral equation for $\eta\left({\bf R}\right)$ is derived via  the approach of real-time propagator, while here we use the approach of imaginary-time propagator.

As shown in Eq. (\ref{eta}), $\eta\left({\bf R}\right)$ is determined by the behavior of $\psi^{\left(+\right)}\left({\bf r}_{A},{\bf r}_{B}\right)$ in the short-range limit ${\bf r}_{A}\rightarrow{\bf r}_{B}={\bf R}$.
Firstly, we notice that Eq. (\ref{eq:LPEquation}) of $\psi^{\left(+\right)}\left({\bf r}_{A},{\bf r}_{B}\right)$ can be re-written as
\begin{align}
\psi^{\left(+\right)}\left({\bf r}_{A},{\bf r}_{B}\right)= & \psi^{\left(0\right)}\left({\bf r}_{A},{\bf r}_{B}\right)+\frac{2\pi a_{s}}{\mu_{AB}}\eta\left({\bf R}\right)\int d{\bf R}'G_{0}\left({\bf r}_{A},{\bf r}_{B};{\bf R}',{\bf R}'\right)\nonumber \\
 & +\frac{2\pi a_{s}}{\mu_{AB}}\int d{\bf R}'G_{0}\left({\bf r}_{A},{\bf r}_{B};{\bf R}',{\bf R}'\right)\left(\eta\left({\bf R}'\right)-\eta\left({\bf R}\right)\right),\label{eq:LPEquation_1}
\end{align}
where
\begin{equation}
G_{0}\left({\bf r}_{A},{\bf r}_{B};{\bf R'},{\bf R}'\right)=\langle{\bf r}_{A},{\bf r}_{B}\vert\frac{1}{E_0+i0^{+}-H_{0}}\vert{\bf r}'_{A},{\bf r}'_{B}\rangle=-\int_{0}^{\infty}d\beta \langle{\bf r}_{A},{\bf r}_{B}\vert e^{\beta (E_0-H_{0})}\vert{\bf R}',{\bf R}'\rangle
\label{eq:G0_3}
\end{equation}
with $E_0=3\omega/2$ as defined in Sec. II. For convenience, we define a function $K_{\beta}\left({\bf r}_{A},{\bf r}_{B};{\bf R}',{\bf R}'\right)$ as
\begin{align}
K_{\beta}\left({\bf r}_{A},{\bf r}_{B};{\bf R}',{\bf R}'\right)\equiv & \langle{\bf r}_{A},{\bf r}_{B}\vert e^{-\beta H_{0}}\vert{\bf R}',{\bf R}'\rangle\nonumber \\
= & \left(\frac{m_{A}m_{B}\omega^{2}}{4\pi^{2}\left(\omega\beta\right)\sinh\left(\omega\beta\right)}\right)^{3/2}\exp\left\{ -\frac{m_{A}}{2\beta}\left({\bf r}_{A}-{\bf R}'\right)^{2}\right\} \nonumber \\
 & \times\exp\left\{ -\frac{m_{B}\omega}{2\sinh\left(\omega\beta\right)}\left[\cosh\left(\omega\beta\right)\left({\bf r}_{B}^{2}+{\bf R}'^{2}\right)-2{\bf r}_{B}\cdot{\bf R}'\right]\right\} .\label{eq:Propagator}
\end{align}
With the help of Eq. (\ref{eq:G0_3}), the second term in the right hand side of Eq. (\ref{eq:LPEquation_1}) can be expressed as
\begin{equation}
\int d{\bf R}'G_{0}\left({\bf r}_{A},{\bf r}_{B};{\bf R}',{\bf R}'\right)=-\int_{0}^{\infty}d\beta F(\beta;{\bf r}_{A},{\bf r}_{B}),
\end{equation}
where
\begin{align}
F(\beta;{\bf r}_{A},{\bf r}_{B}) = &  e^{\beta E_0} \left(\left(\frac{m_{A}m_{B}\omega}{2\pi\left(m_{A}\sinh\left(\omega\beta\right)+m_{B}\left(\omega\beta\right)\cosh\left(\omega\beta\right)\right)}\right)^{3/2}\times\right.\nonumber \\
 & \left.\exp\left\{ -\frac{m_A {\bf r}_{A}^2}{2\beta} -\frac{m_B \omega {\bf r}_{B}^2}{2 \tanh(\omega \beta)}
 +\frac{( m_A \sinh(\omega \beta){\bf r}_{A} + m_B (\omega \beta) {\bf r}_{B} )^2 }{2\beta \sinh(\omega \beta) \left[ m_A + m_B (\omega \beta)\cosh(\omega \beta)\right]} \right\} \right).
\end{align}
We further define
\begin{equation}
\tilde{F}\left(\beta;{\bf r}_{A},{\bf r}_{B}\right) \equiv F(\beta\rightarrow 0;{\bf r}_{A},{\bf r}_{B}) = \left(\frac{m_{A}m_{B}}{2\pi\left(m_{A}+m_{B}\right)\beta}\right)^{3/2}\left.\exp\left\{ -\frac{m_{A}m_{B}}{2\left(m_{A}+m_{B}\right)\beta}\left({\bf r}_{A}-{\bf r}_{B}\right)^{2}\right\} \right).
\end{equation}
Thus, in the limit ${\bf r}_{A}\rightarrow{\bf r}_{B}$, we have
\begin{align}
\int d{\bf R}'G_{0}\left({\bf r}_{A},{\bf r}_{B};{\bf R}',{\bf R}'\right) & =-\int_{0}^{\infty}d\beta \tilde{F} \left(\beta;{\bf r}_{A},{\bf r}_{B}\right)-\int_{0}^{\infty}d\beta\left(F\left(\beta;{\bf r}_{A},{\bf r}_{B}\right)-\tilde{F}\left(\beta;{\bf r}_{A},{\bf r}_{B}\right)\right)\nonumber \\
 & =-\frac{m_{A}m_{B}}{2\pi\left(m_{A}+m_{B}\right)\left|{\bf r}_{A}-{\bf r}_{B}\right|}+F_{1}\left({\bf R}\right)+o\left(\left|{\bf r}_{A}-{\bf r}_{B}\right|\right),\label{eq:F1}
\end{align}
where $F_{1}\left({\bf R}\right)$ is defined as
\begin{equation}
F_{1}\left({\bf R}\right)\equiv-\frac{1}{a_{{\rm trap}}^{3}}\left(\frac{1}{2\pi}\right)^{3/2}\left\{ \int_{0}^{\infty}dx\left(\frac{e^{\frac{3}{2}x}}{\left(\sinh x+\frac{x}{\alpha}\cosh x\right)^{3/2}}e^{-W_{0}\frac{{\bf R}^{2}}{a_{{\rm trap}}^{2}}}-\left(\frac{1}{\left(1+\frac{1}{\alpha}\right)x}\right)^{3/2}\right)\right\}
\end{equation}
with $x=\beta \omega, \alpha=m_{A}/m_{B}$ and $W_{0}=\frac{\left(\cosh x-1\right)+\frac{x}{2\alpha}\sinh x}{\sinh x+\frac{x}{\alpha}\cosh x}$.

Furthermore, in the limit ${\bf r}_{A}\rightarrow{\bf r}_{B}={\bf R}$ the third term of Eq. (\ref{eq:LPEquation_1}) does not diverge, and can be expressed as
\begin{equation}
\int d{\bf R}'G_{0}\left({\bf r}_{A},{\bf r}_{B};{\bf R}',{\bf R}'\right)\left(\eta\left({\bf R}'\right)-\eta\left({\bf R}\right)\right)=\int d{\bf R}'F_{2}\left({\bf R},{\bf R}'\right)\left(\eta\left({\bf R}'\right)-\eta\left({\bf R}\right)\right)+o\left(\left|{\bf r}_{A}-{\bf r}_{B}\right|\right)\label{eq:F2}
\end{equation}
with
\begin{equation}
F_{2}\left({\bf R},{\bf R}'\right)\equiv-\int_{0}^{\infty}d\beta e^{\beta E} K_{\beta}\left({\bf R},{\bf R};{\bf R}',{\bf R}'\right).
\end{equation}
One should notice that $F_{2}\left({\bf R},{\bf R}'\right)$ diverge
when ${\bf R}\rightarrow{\bf R}'$, while $F_{2}\left({\bf R},{\bf R}'\right)\left(\eta\left({\bf R}'\right)-\eta\left({\bf R}\right)\right)$
do not.

In summary, Eq. (\ref{eq:F1}) and Eq. (\ref{eq:F2}) gives the behaviors of the second and third term of Eq. (\ref{eq:LPEquation_1}) in the short-range limit ${\bf r}_{A}\rightarrow{\bf r}_{B}={\bf R}$.
Substituting these results into Eq. (\ref{eq:LPEquation_1}) and Eq. (\ref{eta}), we obtain the integral equation for
$\eta\left({\bf R}\right)$:
 \begin{equation}
\eta\left({\bf R}\right)=\eta_{0}\left({\bf R}\right)+\hat{O}[\eta\left({\bf R}\right)],\label{eqq}
\end{equation}
where $\eta_{0}\left({\bf R}\right)=\psi^{(0)}({\bf R},{\bf R})=\frac{1}{(2\pi)^{\frac{3}{2}}}\phi_0(\bf R)$ is the regularized incident wave function, and the operator $\hat{O}$ is defined as
 \begin{equation}
\hat{O}[\eta\left({\bf R}\right)]=\frac{2\pi a_{s}}{\mu_{AB}}\eta\left({\bf R}\right)F_{1}\left({\bf R}\right)+\frac{2\pi a_{s}}{\mu_{AB}}\int d{\bf R}'F_{2}\left({\bf R},{\bf R}'\right)\left(\eta\left({\bf R}'\right)-\eta\left({\bf R}\right)\right).
\end{equation}
This is Eq. (\ref{ie}) of Sec. II.

In the end of this appendix, we emphasis that
since the angular momentum of this system conserved, the
equation do not have angular dependence, i.e $\eta\left({\bf R}\right)=\eta\left(R\right)$.
Then the equation (\ref{eqq}) of $\eta\left({\bf R}\right)$ can be further reduce to
\begin{equation}
\eta\left(R\right)=\frac{1}{(2\pi)^{\frac{3}{2}}}\phi_0(R)+\frac{2\pi a_{s}}{\mu_{AB}}\eta\left(R\right)F_{1}\left(R\right)+\frac{2\pi a_{s}}{\mu_{AB}}\int dR'R'^{2}F_{2}\left(R,R'\right)\left(\eta\left(R'\right)-\eta\left(R\right)\right),\label{eq:IE}
\end{equation}
where we use $\phi_0({\bf R})= \phi_0(R), F_{1}\left({\bf R}\right)=F_{1}\left(R\right)$ and $F_{2}\left(R,R'\right)$ is
\begin{align}
F_{2}\left(R,R'\right) & =-\int_{0}^{\infty}d\beta e^{\beta E} \left(\int_{0}^{\pi}\int_{0}^{2\pi}\sin\theta d\theta d\phi K_{\beta}\left({\bf R},{\bf R};{\bf R}',{\bf R}'\right)\right)\\
 & =-\frac{1}{a_{{\rm trap}}^{6}}\alpha^{3/2}\left\{ \int_{0}^{\infty}dx\frac{e^{-W_{1}\frac{R'^{2}+R^{2}}{a_{{\rm trap}}^{2}}}}{\left(x\sinh xe^{-x}\right)^{3/2}}\frac{\sinh\left(2W_{2}\,\frac{RR'}{a_{{\rm trap}}^{2}}\right)}{\left(2\pi\right)^{2}W_{2}}\right\}
\end{align}
with $x=\beta \omega, W_{1}=\frac{\alpha}{2x}+\frac{1}{2\tanh x},W_{2}=\frac{\alpha}{2x}+\frac{1}{2\sinh x}$.
Now both $F_{1}\left(R\right)$ and $F_{2}\left(R,R'\right)$
can be calculated numerically. Solving the integral equation Eq. (\ref{eq:IE})
will give the numerical solution of $\eta\left(R\right)$.
\end{widetext}

\end{document}